\documentclass{article}
\usepackage{multirow}
\usepackage{graphicx}
\usepackage{amssymb}
\usepackage{caption}
\usepackage{subcaption}
\usepackage[left=1.5cm,right=1.5cm,top=1.5cm,bottom=1.5cm]{geometry}
\usepackage{authblk}
\begin{document}
\title{Exploring density dependent B as a suitable parameter in higher dimensional approach with a non-linear equation of state}
\author[1]{Koushik Ballav Goswami\thanks{koushik.kbg@gmail.com}}
\author[2]{Debadri Bhattacharjee\thanks{debadriwork@gmail.com}}
\author[3]{Anirban Saha\thanks{anirban.astro9@gmail.com}}
\author[4]{Pradip Kumar Chattopadhyay\thanks{pkc$_{-}76$@rediffmail.com}}
\affil[1,2,4]{IUCAA Centre for Astronomy Research and Development (ICARD), Department of Physics, Cooch Behar Panchanan Barma University, Vivekananda Street, District: Cooch Behar, \\ Pin: 736101, West Bengal, India}
\affil[3]{Department of Physics, Alipurduar University, Alipurduar, Pin-736122, West Bengal, India.}
\maketitle
\begin{abstract}
	In this investigation, we present a singularity free interior solution of the Einstein field equation for a class of anisotropic compact objects in dimensions $D\geq4$. In accordance with the concept of Vaidya and Tikekar, the geometry of the physical $(D-1)$-space of a star corresponding to $t=constant$ hypersurface is assumed to be of a $(D-1)$ spheroid. For the fulfilment of causality condition, a limit of the spheroidal parameter ($\lambda$) is noted depending on the values of amount of anisotropy ($\alpha$) and space-time dimensions ($D$). We note that by switching off the extra parameters ($\alpha$ and $D$), previously obtained limit of $\lambda$ can be generated. To validate our findings, we compare the results obtained from our model with observational data of PSR J1614-2230 (mass=$1.908^{+0.016}_{-0.016}M_{\odot}$, radius=$11.93^{+0.50}_{-0.50}km$). It is noted that the best fit equation of state corresponds to polynomial equation of state of the order of five. We use this finding to develop a density dependent MIT bag model which seems to be useful for the correct description of compact object in our model. The mass radius relation shows that our model mimics a wide range of recently observed pulsars in four and higher dimensions. Furthermore, we also found that our model exhibits stability according to Generalised TOV equation, Herrera cracking condition, and the adiabatic index.
\end{abstract}
\section*{Keyword}
	Higher dimensions ; Pressure anisotropy ; Strange quark star ; MIT bag model ; Density dependent bag parameter
\section{Introduction} \label{intro}
In the year 1964, Gell-Mann \cite{GellMann} and Zweig \cite{Zweig} independently proposed that hadrons are made up of more fundamental particles known as quarks, a theory that later received experimental validation. On the theoretical context, the quark matter hypothesis introduced by Bodmer \cite{Bodmer} and Witten \cite{Witten} led many researchers to explore a completely new category of compact astrophysical objects made of Strange Quark Matter (SQM), known as `Strange Stars' (SS) \cite{Madsen,Baym,Glendenning,Alcock}. After the seminal work of Bodmer and Witten, the study of various aspects of quark matter has garnered significant interest from both astrophysicists and particle physicists. Witten suggested that at the critical temperature $T_{c}\sim 200 MeV$, hadrons break up into a small fraction of coloured objects, such as quarks and gluons, leading to the formation of coloured particles known as quark nuggets (QNs) which consist of $u$, $d$ and $s$ quarks. Later, a large group of researchers investigated the properties of SQM \cite{Weber,Applegate,Farhi,Chandra}. Bhattacharyya and his collaborators \cite{Bhattacharyya1,Bhattacharyya2,Bhattacharyya3} suggested that shortly after the Big Bang, within a few microseconds, the universe experienced a quark-gluon phase transition, potentially marking the origin and persistence of quark matter. However, the exact nature of the confining force responsible for triggering this phase transition remains largely unknown. This ultimately makes the Equation of State (EoS) for SQM to be uncertain. Therefore, to study SS, various phenomenological models of quark confinement came into existence. One of such quark confinement models was the MIT bag model \cite{Chodos}. Earlier, it was a challenging task for physicists to think about free quarks outside hadrons. In Quantum Chromodynamics (QCD), vacuum is characterised by the property that it cannot support the propagation of quarks and gluon fields. In such vacuum, different phase may coexist within small domains. Such domains are like 'bubble's having the characteristic size of a hadron and allows free propagation of quarks and gluons. In MIT bag model, hadrons are modeled as spherical bags, with quarks treated as Dirac particles that are permanently confined within the finite volume of the bag. The radius of this bag corresponds to the size of the hadron. In this model, a Lorentz scalar term called the bag parameter ($B$) is introduced to compensate the internal pressure exerted by the freely moving quarks and gluons. A strange property of the bag is that the quarks and gluons become heavy as they tend to move towards the surface and hence cannot escape to the normal phase of the vacuum. Assuming quark species to be massless and noninteracting, the pressure exerted by individual quark species is $p_{q}=\rho_{q}/3$, where $\rho_{q}$ is the energy density of the $q^{th}$ species. For quark confinement, the total external pressure ($p$) and energy density ($\rho$) are $p=p_{q}-B$ and $\rho=\rho_{q}+B$ respectively \cite{Kettner,Kapusta}. Consequently, the EoS in MIT bag model can be obtained in the form:
\begin{equation}
	p=\frac{1}{3}(\rho-4B). \label{eq0}
\end{equation}
In case of zero external pressure ($p=0$), quark matter consists with only $u$ and $d$ quarks is likely unstable \cite{Madsen}. As such two flavour quark matter is concerned, the energy per baryon ($E_{B}$) of such system is $934$ $B^{1/4}_{145}~MeV$ where, $B^{1/4}_{145}=\frac{B^{1/4}}{145}$. Inclusion an extra flavour ($s$) into the system effectively raises the energy per baryon to a level $829$ $B^{1/4}_{145}~MeV$. Thus in case of three flavour quark matter, the system gains approximately 100 $MeV$ as energy per baryon and as a result the extra fermi well allows much denser packing and thereby increases the overall stability of the system. Physical behaviour of SS for a MIT bag model type EoS has been reported in refs.\cite{Brilenkov,Paulucci,Arbanil,Lugones,Chowdhury}. Kalam et al. \cite{Kalam} presented an analytical model of SS based on MIT bag EoS and showed that bag parameter $B$ can take value as high as $203~MeV fm^{-3}$. Aziz et al. \cite{Aziz} on the other hand, estimated a wide range of $B$ within $41.58~MeV fm^{-3}$ to $319.31 MeV fm^{-3}$. Although in the phenomenological bag model the bag parameter $B$ is considered as a constant, theoretical research \cite{Reinhardt} unveils that at high baryon number density which prevails at the core of SS, it is more practical to treat $B$ as a density dependent quantity. Using this concept a vast number of work \cite{Chakrabarty1,Chakrabarty2,Chakrabarty3,Chakrabarty4,Peng,Prasad,Zhu} has been done in this field of research.\par
There are several physical processes such as exotic phase transitions \cite{Collins,Itoh,Migdal,Sawyer,Sokolov} especially to a pion condensate state \cite{Hartle}, existence of a solid core \cite{Ruderman,Cameron}, mixture of two different gases \cite{Letelier} etc. could develop pressure anisotropy in self-gravitating objects. Herrera \cite{Herrera} argued that systems that are initially assumed to have isotropic pressure can develop pressure anisotropy as a result of dynamic processes. This implies that the final equilibrium configuration emerging from such a dynamic phase may preserve the anisotropic pressure formed during its evolution. Therefore, a system that begins with isotropic pressure could still display anisotropic pressure in its ultimate equilibrium state. Cosenza et al. \cite{Cosenza} investigated a heuristic method for deriving interior solutions to Einstein Field Equations (EFEs) for anisotropic matter, starting from existing solutions formulated for isotropic matter distributions. For a super-dense compact object in which density exceeds several times that corresponds to nuclear drip, development of pressure anisotropy is inevitable \cite{Ruderman2,Canuto}. Apart from all such factors, the presence of superfluid of type 3A is considered to be one of the prime factors creating anisotropy in NS or SS \cite{Kippenhahn}. Quantum mechanically, Pauli's exclusion principle forbids the accommodation of nucleon in a single energy state. At nuclear scale, strong interaction comes into play which can withstand a star against strong gravitational pull. However, the scenario changes at low temperatures when nucleon form Cooper pairs \cite{Broglia}. Such Cooper pair are by virtue bosons and behave collectively. This collective behaviour of nucleon bear a resemblance to nonviscous nucleonic condensate similar to He-3 superfluid. Inside of NS or SS, high pressure environment effectively raises the critical temperature for Cooper pair formation allowing the formation of Cooper pairs even under the existence of such extreme condition. It is hypothesised that three different kinds of superfluid may exist inside a NS or SS \cite{Page}. In an effort to provide a more comprehensive framework for modeling compact stars, Bowers and Liang \cite{Bowers} incorporated the concept of pressure anisotropy into their analysis. Their work explored how anisotropy influences various stellar properties, such as the degree of compactness and the relationship between mass and radius. A general algorithm for generating static anisotropic solutions by Lake \cite{Lake} has led many researchers to work in this field of research. Baskey et al. \cite{Baskey} has studied the slow rotation effects of anisotropic compact objects. Recently Pradhan et al. \cite{Pradhan} has developed an anisotropic model of geometrically deformed strange stars in the context of teleparallel gravity. A general singularity free solution of the EFE for anisotropic fluid has been investigated by Baskey et al. \cite{Baskey2}. Anisotropic extension of Durgapal-Fuloria solution has been obtained by Maurya et al. \cite{Maurya}. Pandya et al. \cite{Pandya} obtained a global regular solution of the Finch-Skea ansatz in anisotropic domain.\par
In string theory and its brane world extensions \cite{Green,Polchinski}, it is well established that more than four spacetime dimensions are required to consistently describe the fundamental structure of the Universe. This higher-dimensional framework is essential for formulating a unified and consistent quantum theory that incorporates all fundamental forces of nature, including gravity. The concept of extra dimensions was first initiated by Kaluza and Klein \cite{Kaluza,Klein} who tried to unify gravity with fundamental forces in nature. Chodos and Detweiler \cite{Chodos2} pointed out that the early universe was might be in a higher dimensional stage which during its course of evolution shrank into a four dimensional space. Due to dynamical contraction, the extra dimensions became so minute that its effects are hardly to be detectable. Later, various astrophysical models such as, Schwarzschild and Reissner-Nordstr{\"o}m black holes \cite{Chodos3}, rotating Kerr black hole \cite{Myers,Mazur,Frolov,Dianyan}, no hair theorem \cite{Sokolowski}, black holes in compactified spacetime \cite{Myers2}, theory of Hawking radiation \cite{Myers} etc. were generalised in higher dimensional framework. Recently, Zahra et al. \cite{Zahra} have developed a general framework to study the gravitational collapse of compact objects. In their analysis, they have employed a non-comoving coordinate system which resembles a five-dimensional Schwarzschild metric and thus by matching with the exterior Vaidya metric some valuable insights of gravitational collapse have been found. In 1982, Vaidya and Tikekar \cite{Tikekar} presented a model describing super-dense stellar structure. In their formalism, the physical 3-space corresponding to $t=constant$ hypersurface of a star was assumed to take the geometry of a spheroid. The relativistic solution of the EFE for such consideration allows density of the order of $2\times10^{14} gm/cm^3$ and thus capable of describing super-dense stellar configuration. Later Paul et al. \cite{Paul} generalized the Vaidya-Tikekar solution in higher dimensional domain for anisotropic star. Vaidya-Tikekar solution has been employed by a vast group of authors for the study of SS \cite{Goswami,Saha,Goswami2,Sharma,Hansraj,Sharif,Nag,Goswami3}.\par
Our present study involves solution of EFE, representing anisotropic cold compact stars through the choice of $g_{rr}$ metric function according to Vaidya and Tikekar \cite{Tikekar} in the framework of General Relativity. Using the solution, we try to explore the EoS of a chosen compact star and relate the result with MIT bag model EoS and hence try to analyse the physical characteristics of the stellar model.\par
The structure of the paper is organised as follows: In Sec.~\ref{efe}, solution of the Einstein field equation in presence of extra dimensions ($D\geq4$) has been presented for Vaidya-Tikekar metric ansatz by taking anisotropy in pressure into account. Sec.~\ref{cond} states various criteria required for a physically admissible stellar solution. In Sec.~\ref{match}, we have shown the boundary conditions which are essential to be applied at the stellar surface. Sec.~\ref{sound} is devoted to constrain the model parameters based on the causality limit. Physical application of the obtained stellar solution has been presented in Sec.~\ref{physical} for a known compact object using its observed mass and radius. Various energy conditions and the equation of state have also been discussed in this section. Relation between mass and radius of compact objects have been illustrated in Sec.~\ref{mass}. Sec.~\ref{stab} reflects various stability criteria needed for a stable stellar structure in hydrostatic equilibrium. Finally, we conclude on our obtained model by discussing some of its important features in Sec.~\ref{conc}.

\section{Exact solution of EFEs in higher dimensions with anisotropic pressure}\label{efe}
To arrive at the field equations, we assume that the space-time geometry of a compact object in a $D$ dimension manifold is spherically symmetric and described by the following line element:
\begin{equation}
	ds^2=-e^{2\nu(r)}dt^2+e^{2\mu(r)}dr^2+r^2d{\Omega_{n}}^2, \label{eq1}
\end{equation}
where we have considered $n=D-2$ and the term $d{\Omega_{n}}^2=d{\theta_{1}}^2+sin^2\theta_{1}d{\theta_{2}}^2+sin^2\theta_{1}sin^2\theta_{2}d{\theta_{3}}^2 +........+(sin^2\theta_{1}sin^2\theta_{2}........sin^2\theta_{n-1})d{\theta_{n}}^2$ represents the angular part of the metric on $n$-sphere. For a fluid sphere having anisotropic pressures, the energy momentum tensor describing the interior matter content is expressed as follows:
\begin{equation}
	T^{i}_{j}=\mbox{diag}~(-\rho, p_{r}, p_{t}, p_{t},........., p_{t}). \label{eq2}
\end{equation}
The tensorial form of Einstein field equations is given by
\begin{equation}
	{\bf R}^{i}_{j}-\frac{1}{2}\delta^{i}_{j}{\bf R}=8\pi G_{D}T^{i}_{j}, \label{eq3}
\end{equation}
here, the four dimensional Newtonian gravitational constant ($G$) has been replaced by its higher dimensional counterpart $G_{D}=GV_{D-4}$, where $V_{D-4}$ represents the volume of the extra dimensions and is related to the radius of a compact object through the relation $V_{D-4}=\frac{\pi^{(D-4)/2}b^(D-4)}{\Gamma(1+(D-4)/2)}$.
Using Eqs.~(\ref{eq1}) and ~(\ref{eq2}) the field equations ~(\ref{eq3}) gives the following set of equations:
\begin{equation}
	\frac{ne^{-2\mu}\mu^{\prime}}{r}+\frac{n(n-1)}{2r^2}(1-e^{-2\mu})=8\pi G_{D}\rho. \label{eq4}
\end{equation}
\begin{equation}
	\frac{ne^{-2\mu}\nu^{\prime}}{r}-\frac{n(n-1)}{2r^2}(1-e^{-2\mu})=8\pi G_{D}p_{r}. \label{eq5}
\end{equation}
\[
e^{-2\mu}\left[\nu^{\prime\prime}+{\nu^{\prime}}^2-\mu^{\prime}\nu^{\prime}-\frac{(n-1)(\mu^{\prime}-\nu^{\prime})}{r}\right]
\]
\begin{equation}
	\;\;\;\;-\frac{(n-1)(n-2)}{2r^2}(1-e^{-2\mu})=8\pi G_{D}p_{t}. \label{eq6}
\end{equation}
To solve the set of Eqs.~(\ref{eq4}) - (\ref{eq6}), we assume the ansatz proposed by Vaidya and Tikekar \cite{Tikekar}:
\begin{equation}
	e^{2\mu}=\frac{1+\lambda\frac{r^2}{\mathcal{R}^2}}{1-\frac{r^2}{\mathcal{R}^2}}, \label{eq7}
\end{equation}
In this approach, the $t =$ constant section of space-time is equivalent to the geometry of $(D-1)$ ($\equiv n+1$) spheroid immersed in $D$ dimensional Euclidean space. The parameter $\lambda$ defines the spheroidicity of the ($n+1$) spheroid and $\mathcal{R}$ is related to the curvature of the hypersurface. To simplify Eqs.~(\ref{eq5}) and (\ref{eq6}), we use the transformation $x^2=1-\frac{r^2}{\mathcal{R}^2}$, $\psi=e^{\nu(r)}$ and consider $\Delta=p_{t}-p_{r}$ to be the measure of pressure anisotropy \cite{Mak,Dev,Chaisi,Thomas,Maharaj}. Now subtracting equation~(\ref{eq5}) from equation~(\ref{eq6}) we obtain the following second order differential equation in $x$:
\[
\left[1+\lambda(1-x^2)\right]\psi_{xx}+ax\psi_{x}+\lambda(\lambda+1)(n-1)\psi
\]
\begin{equation}
	\;\;\; -8\pi G_{D}\Delta \mathcal{R}^2\frac{[1+\lambda(1-x^2)]^2}{1-x^2}\psi=0. \label{eq8}
\end{equation}
Equation~(\ref{eq8}) can not be integrated unless a choice for the pressure anisotropy $\Delta$ is made. In this article, we choose $\Delta$ as follows:
\begin{equation}
	\Delta=\frac{\alpha \lambda^{2}(1-x^2)(n-1)}{8\pi G_{D}\mathcal{R}^2[1+\lambda(1-x^2)]^2}. \label{eq9}
\end{equation}
The above choice for $\Delta$ ensures the regularity of anisotropic force ($\frac{n\Delta}{r}$) at the centre of a star. Similar form of $\Delta$ was also taken by Goswami et al. \cite{Goswami} to construct anisotropic stellar model in higher dimensions. To simplify equation~(\ref{eq9}) further, we use the transformation $z=\sqrt{\lambda/(\lambda+1)}x$ and finally get the following second order differential equation in $z$:
\begin{equation}
	(1-z^2)\psi_{zz}+z\psi_{z}+(n-1)(1+\Omega)\psi=0. \label{eq10}
\end{equation}
In the above equation, $\Omega=\lambda(1-\alpha)$ is a constant. Following the procedure adopted by Mukherjee et al. \cite{Mukherjee}, the general solution of equation~(\ref{eq10}) is obtained as 
\begin{equation}
	\psi=A\left[\frac{\cos{[(\omega+1)\eta+\phi]}}{\omega+1}-\frac{\cos{[(\omega-1)\eta+\phi]}}{\omega-1}\right], \label{eq11}
\end{equation}
where $\omega=\sqrt{n+(n-1)\Omega}$ and $\eta=cos^{-1}z$. The unknown constants $A$ and $\phi$ can be determined from the matching conditions at  the boundary of a star. It is to be mentioned that for a realistic stellar model the value of the constant $\phi$ should be real, otherwise, the metric potential $\nu(r)$ will be imaginary which is not practical. Since, both the metric potentials ($\mu$ and $\nu$) have been determined, we can now obtain the expressions for the energy density ($\rho$) and pressures ($p_{r}$, $p_{t}$) using Eqs.~(\ref{eq4})-(\ref{eq6}) as given below:
\begin{equation}
	\tilde{\rho}=8\pi G_{D}\rho=\frac{n}{2\mathcal{R}^2(1-z^2)}\left[n-1+\frac{2}{(1+\lambda)(1-z^2)}\right], \label{eq12}
\end{equation}
\begin{equation}
	\tilde{p_{r}}=8\pi G_{D}p_{r}=-\frac{1}{\mathcal{R}^2(1-z^2)}\left[\frac{n(n-1)}{2}+\frac{nz}{\lambda+1}\left(\frac{\psi_{z}}{\psi}\right)\right], \label{eq13}
\end{equation}
\begin{equation}
	\tilde{\Delta}=8\pi G_{D}\Delta=\frac{(n-1)\alpha \lambda(\lambda-(\lambda+1)z^2)}{\mathcal{R}^2(1-z^2)^2(\lambda+1)^2}, \label{eq14}
\end{equation}
\begin{equation}
	\tilde{p_{t}}=\tilde{p_{r}}+\tilde{\Delta}. \label{eq15}
\end{equation}
In Eqs.~(\ref{eq12})-(\ref{eq15}), we have rescaled the equations in term of a factor $8\pi G_{D}$ so that the units of energy density and pressures can be obtained in geometrized unit ($km^{-2}$). The total mass contained within radius $'b'$ of a compact object in $D$ dimensions is given by
\begin{equation}
	M=A_{n}\int_{0}^{b}r^{n}\tilde{\rho}(r)dr \label{eq16}
\end{equation}
where $A_{n}=\frac{2{\pi}^{\frac{n+1}{2}}}{\Gamma(\frac{n+1}{2})}$. Using Eq.~(\ref{eq12}), we can obtain the total mass of a compact object of radius $'b'$ by numerically integrating Eq.~(\ref{eq16}).
\section{Conditions of a physically acceptable stellar solution}\label{cond}
One of the possible ways to obtain a mathematical model of compact objects is to finding exact solution of the Einstein field equation~(\ref{eq3}). For this purpose, one has to choose one of the metric potentials as an ansatz. There are several metric ansatz have been developed till date which immediately satisfies the field equations. However, thorough investigations show that not all of them are physically acceptable as pointed out by Delgaty and Lake \cite{Delgaty}. Therefore, after obtaining relativistic solution of a compact object, it is essential to verify the physical acceptability of the solution. In the case of an isolated static spherically symmetric perfect fluid sphere, the following conditions need to be verified:
\begin{itemize}
	\item The metric functions $e^{2\nu(r)}$ and $e^{2\mu(r)}$ should be regular and singularity free.
	\item The physical parameters like energy density ($\rho$), pressures ($p_{r}$, $p_{t}$) and $\Delta$ must be well behaved and non-singular throughout the interior of a star.
	\item Energy density ($\rho$), radial ($p_{r}$), and tangential ($p_{t}$) pressures should be positive in the interior of stellar model. The radial pressure ($p_{r}$) must vanish at the boundary of a star ($p_{r}(b)=0$). However, the tangential pressure may be non-vanishing at the boundary ($p_{t}(b)\neq0$).
	\item At the centre of a star, both radial and tangential pressures become equal indicating that $\Delta=0$ at the centre of a star.
	\item Energy density and pressures should be monotonically decreasing function of radial coordinate $'r'$ i.e. $\frac{d\rho}{dr}<0$, $\frac{dp_{r}}{dr}<0$, and $\frac{dp_{t}}{dr}<0$.
	\item The sound velocities inside a stable compact object should not be superluminal which requires the fulfilment of causality conditions i.e. $v_{r}^{2}=\frac{dp_{r}}{d\rho}<1$, and $v_{t}^{2}=\frac{dp_{t}}{d\rho}<1$.
	\item The Zeldovich condition \cite{Zeldovich} should be satisfied which states that the ratio of the radial pressure and density at the centre of a stable stellar structure should always be less than or equal to 1 (i.e. $\frac{p_{r}(0)}{\rho(0)}\leq1$). This condition ensures that pressure is not too high relative to density of a star which otherwise would lead to instability.
	\item The interior metric functions should match smoothly with the exterior vacuum Schwarzschild metric at the boundary of a star. 
\end{itemize}
\section{Matching conditions}\label{match}
\begin{itemize}
	\item For the regularity of the metric potentials, the interior solution should be smoothly matched with the exterior vacuum Schwarzschild solution. The vacuum space-time geometry is best described by a space-time metric given by \cite{Shen}:
	\begin{equation}
		ds^2=-\left(1-\frac{K_{n}}{r^{n-1}}\right)dt^2+\left(1-\frac{K_{n}}{r^{n-1}}\right)^{-1}dr^2+r^2d\Omega_{n}^{2}, \label{eq17}
	\end{equation}
	where, the dimension dependent constant $K_{n}$ is related to the mass ($M$) of a compact object given by:
	\begin{equation}
		M=\frac{nA_{n}K_{n}}{16\pi G_{D}}, \label{eq18}
	\end{equation}  
	where, $A_{n}=\frac{2\pi^{\frac{n+1}{2}}}{\Gamma(\frac{n+1}{2})}$. Matching of the interior and exterior metric potential gives:
	\begin{equation}
		e^{2\mu(r=b)}=\left(1-\frac{K_{n}}{b^{n-1}}\right)^{-1}. \label{eq19}
	\end{equation}
	Equation~(\ref{eq19}) can be employed to determining the parameter $\mathcal{R}$. Alternatively, $\mathcal{R}$ can also be determined from the knowledge of the central density ($\rho_{c}$) of a star. From equation~(\ref{eq12}), we note that the central density of a star is given by
	\begin{equation}
		\tilde{\rho_{c}}=\frac{n(n+1)(\lambda+1)}{2\mathcal{R}^2}. \label{eq20}
	\end{equation}
	Equation~(\ref{eq20}) shows that central density of a star is a dimension dependent quantity and increases rapidly with an increase of the space-time dimension.
	\item The boundary of a star is defined as the surface at which the radial pressure vanishes. Thus, if $b$ be the radius of a star, the condition $p_{r}(b)=0$ should be satisfied. Therefore, equation~(\ref{eq13}) gives the following condition
	\begin{equation}
		\frac{\psi_{z}(z_{b})}{\psi(z_{b})}=-\frac{(n-1)(\lambda+1)}{2z_{b}}, \label{eq21}
	\end{equation}
	where, $z_{b}=\sqrt{\frac{\lambda}{\lambda+1}}\sqrt{1-\frac{b^2}{\mathcal{R}^2}}$. Again, differentiating equation~(\ref{eq11}) with respect to $z$ and evaluating at $r=b$ we get
	\begin{equation}
		\frac{\psi_{z}(z_{b})}{\psi(z_{b})}=\frac{\omega^2-1}{\sqrt{1-z_{b}^2}}\left[\frac{sin\{(\omega+1)\eta+\phi\}-sin\{(\omega-1)\eta+\phi\}}{(\omega-1)cos\{(\omega+1)\eta+\phi\}-(\omega+1)cos\{(\omega-1)\eta+\phi\}}\right]. \label{eq22}
	\end{equation}
	Now, equating Eqs.~(\ref{eq21}) and (\ref{eq22}), we get an expression of the constant $\phi$ as follows
	\begin{equation}
		tan(\phi)=\frac{\tau cot(\eta_{b})-tan(\omega \eta_{b})}{1+\tau cot(\eta_{b})tan(\omega \eta_{b})}, \label{eq23}
	\end{equation}
	where, $\tau=\frac{1}{\omega}\left[\frac{2(\omega^2-1)}{(n-1)(\lambda+1)}-1\right]$ and $\eta_{b}=cos^{-1}z_{b}$.
\end{itemize}
\section{Sound velocities and bounds on the model parameters}\label{sound}
The square of the radial sound velocity is obtained from the slope of the EoS i.e. $\frac{dp_{r}}{d\rho}$ and given by
\begin{equation}
	v_{r}^{2}=\frac{dp_{r}}{d\rho}=\frac{z(1-z^2)^2\left(\frac{\psi_{z}}{\psi}\right)^2-(1-z^2)\left(\frac{\psi_{z}}{\psi}\right)-z(1-z^2)(n-1)\lambda \alpha}{4z+z(1-z^2)(n-1)(\lambda+1)}. \label{eq24}
\end{equation}
In a similar manner we can also obtain the expression of the square of tangential sound velocity as follows:
\begin{eqnarray}
	v_{t}^{2}&=&\frac{dp_{t}}{d\rho}=v_{r}^{2}+\frac{d\Delta}{d\rho} \nonumber \\
	&& = v_{r}^{2}-\frac{2(n-1)\alpha \lambda[1-\lambda+z^2(1+\lambda)]}{n(1+\lambda)[4+(n-1)(1+\lambda)(1-z^2)]}. \label{eq25}
\end{eqnarray}
Equation~(\ref{eq25}) shows that $v_{t}^2<v_{r}^2$. Therefore, for determining the bounds on model dependent parameters, it is only essential to consider the restriction on $v_{r}^2$. It has been reported by several authors \cite{Bedaque,Han,Altiparmak,Ecker,Brandes,Braun,Ecker2,Han2} that at intermediate nuclear density, the velocity of sound may exceed the bound $v_{r}^2>\frac{1}{3}$. It has also been suggested by some researchers that matter inside a NS where density excessively high, may be superluminal \cite{Glass,Ruderman3,Ellis}. However, this remains a debatable issue till date. Therefore, in our present work we assume that causality condition ($v_{r}^2<1$) is not violated inside a compact object. Satisfaction of causality condition ($0<v_{r}^2<1$) requires that
\begin{equation}
	\frac{1}{1-z^2}\left(\frac{1}{2z}-\zeta\right)\leq\frac{\psi_{z}}{\psi}\leq\frac{1}{1-z^2}\left(\frac{1}{2z}+\zeta\right), \label{eq26}
\end{equation}
where, $\zeta=\sqrt{4+(n-1)(\lambda+1)(1-z^2)+\frac{1}{4z^2}+\lambda\alpha(n-1)(1-z^2)}$. Again from the condition of non-negative value of radial pressure ($p_{r}\geq0$), equation~(\ref{eq13}) gives
\begin{equation}
	\left(\frac{\psi_{z}}{\psi}\right)\leq-\frac{(n-1)(\lambda+1)}{2z}. \label{eq27}
\end{equation}
Now, combining Eqs.~(\ref{eq26}) and (\ref{eq27}), we get an upper bound on the square of the reduced radius $(\frac{b}{\mathcal{R}})^2$
\begin{equation}
	\left(\frac{b}{\mathcal{R}}\right)^2\leq \frac{(1+\lambda)Y-[6+n(n+1)+8\lambda+2(n-1)(n+\alpha)\lambda+(n-1)[n-2(1+\alpha)]\lambda^2]}{(n-1)\lambda[3+\lambda(2+4\alpha-\lambda)+n(1+\lambda)^2]}, \label{eq28}
\end{equation}
where, $Y=\sqrt{33+14n+17n^2+2(n-1)[7+17n+2(n+7)\alpha]\lambda+(n-1)^2[17+4\alpha(1+\alpha)]\lambda^2}$. Since ($\frac{b^2}{\mathcal{R}^2}$) is real and positive, from equation~(\ref{eq28}), we obtain a lower limit of the spheroidal parameter ($\lambda$)
\begin{equation}
	\lambda>\frac{7+2n-n^2-2\sqrt{9+6n+n^2+\alpha-n\alpha-n^2\alpha+n^3\alpha}}{n^2-4n-13+4\alpha-4n\alpha}. \label{eq29}
\end{equation}
It is interesting to note that the bound given in equation~(\ref{eq29}) reduces to the case $\lambda>\frac{3}{17}$ when $D=4$ and $\alpha=0$ as previously obtained by Mukherjee et al. \cite{Mukherjee}.
\begin{table}[ht!]
	\centering
	\caption{Tabulation of lower limit of lambda ($\lambda_{lower}$) as obtained from equation~(\ref{eq29}).}
	\label{Tab1} 
	\begin{tabular}{ccc}
		\hline
		Dimension & Anisotropy parameter & \multirow{2}{*}{$\lambda_{lower}$}  \\
		$D=n+2$	  &    $\alpha$          &  	                               \\ \hline
		\multirow{5}{*}{4}& 0            & 0.176  \\
		& 0.3          & 0.175  \\
		& 0.5          & 0.173  \\
		& 0.7          & 0.172  \\
		& 0.9          & 0.171  \\ \hline 
		\multirow{5}{*}{5}& 0            & 0.5    \\
		& 0.3          & 0.477  \\
		& 0.5          & 0.463  \\
		& 0.7          & 0.451  \\
		& 0.9          & 0.439  \\ \hline 
		\multirow{5}{*}{6}& 0            & 1.154  \\
		& 0.3          & 1.013  \\
		& 0.5          & 0.943  \\
		& 0.7          & 0.885  \\
		& 0.9          & 0.837  \\ \hline 
		\multirow{5}{*}{7}& 0            & 3      \\
		& 0.3          & 2.130  \\
		& 0.5          & 1.823  \\
		& 0.7          & 1.610  \\
		& 0.9          & 1.452  \\ \hline 
		\multirow{5}{*}{8}& 0            & 35     \\
		& 0.3          & 5.730  \\
		& 0.5          & 3.906  \\
		& 0.7          & 3.035  \\
		& 0.9          & 2.520  \\ \hline 
		\multirow{5}{*}{9}& 0            & $-$    \\
		& 0.3          & $-$    \\
		& 0.5          & 14.810 \\
		& 0.7          & 7.129  \\
		& 0.9          & 4.846  \\ 
		
		\hline                                                                                                                       
	\end{tabular}
\end{table}
In Table~\ref{Tab1}, we have shown the lower bounds of the spheroidal parameter ($\lambda$) which have been calculated from equation~(\ref{eq29}) for various choices of space-time dimension ($D$) and pressure anisotropy parameter ($\alpha$). It can be noted that for lower space-time dimensions (e.g. when $D=4,5$), the bound $\lambda_{lower}$ is not significantly dependent on $\alpha$. However, for $D>5$, the effect of $\alpha$ on the bound of $\lambda$ becomes significant. We also report that when space-time dimension is too high ($D>8$), lower value of $\alpha$ gives negative bounds of $\lambda$ which is generally unphysical. At such high value of $D$, non-negative value of $\lambda_{lower}$ can only be obtained if higher value of $\alpha$ is considered. Therefore, it appears that a compact object becomes more anisotropic in nature when the dimension of the space-time takes larger value.
\section{Physical application of the model}\label{physical}
\begin{figure}[ht!]
	\centering
	\includegraphics[width=0.5\textwidth]{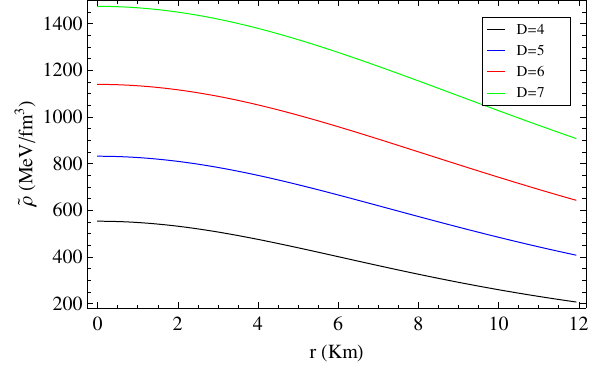}
	\caption{Radial variation of energy density ($\tilde{\rho}$) inside PSR J1614-2230 in different space-time dimensions ($D$). Here, the pressure anisotropy parameter $\alpha=0.3$ for $D=4,5$ and $0.98$ for $D=6,7$ respectively. Value of the spheroidal parameter $\lambda$ is set to 40.}
	\label{Fig1}
\end{figure}

\begin{figure}[ht!]
	\centering
	\includegraphics[width=0.5\textwidth]{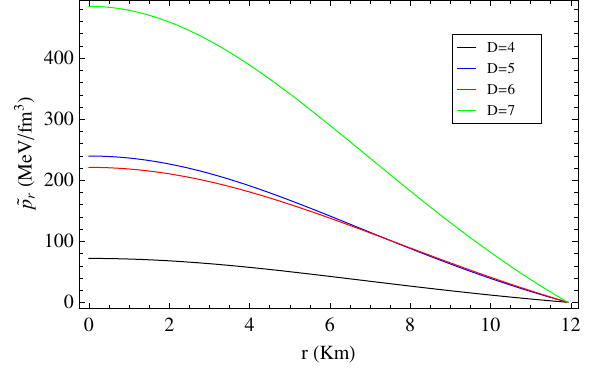}
	\caption{Radial variation of radial pressure ($\tilde{p_{r}}$) inside PSR J1614-2230 in different space-time dimensions ($D$). Here, the pressure anisotropy parameter $\alpha=0.3$ for $D=4,5$ and $0.98$ for $D=6,7$ respectively. Value of the spheroidal parameter $\lambda$ is set to 40.}
	\label{Fig2}
\end{figure}

\begin{figure}[ht!]
	\centering
	\includegraphics[width=0.5\textwidth]{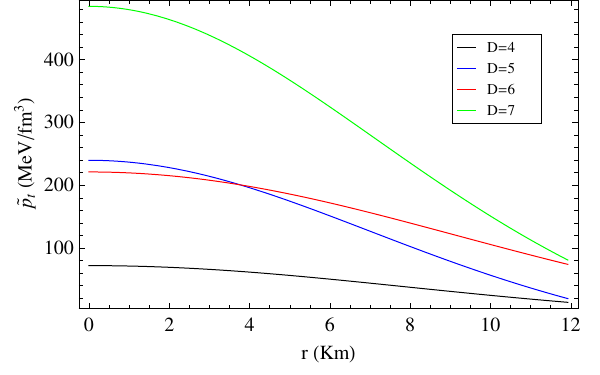}
	\caption{Radial variation of tangential pressure ($\tilde{p_{t}}$) inside PSR J1614-2230 in different space-time dimensions ($D$). Here, the pressure anisotropy parameter $\alpha=0.3$ for $D=4,5$ and $0.98$ for $D=6,7$ respectively. Value of the spheroidal parameter $\lambda$ is set to 40.}
	\label{Fig3}
\end{figure}

\begin{figure}[ht!]
	\centering
	\includegraphics[width=0.5\textwidth]{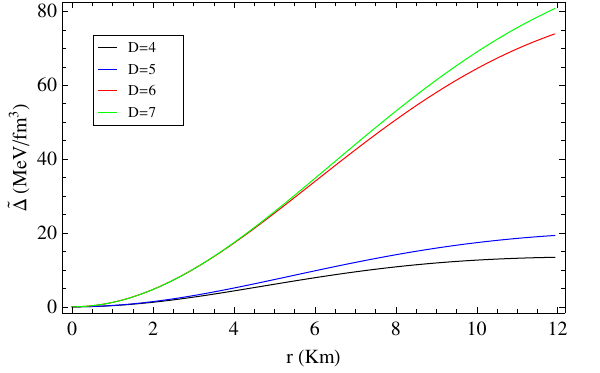}
	\caption{Radial variation of pressure anisotropy ($\tilde{\Delta}$) inside PSR J1614-2230 in different space-time dimensions ($D$). Here, the pressure anisotropy parameter $\alpha=0.3$ for $D=4,5$ and $0.98$ for $D=6,7$ respectively. Value of the spheroidal parameter $\lambda$ is set to 40.}
	\label{Fig4}
\end{figure}

\begin{figure}[t!]
	\centering
	\begin{subfigure}[t]{0.5\textwidth}
		\centering
		\includegraphics[width=1\textwidth]{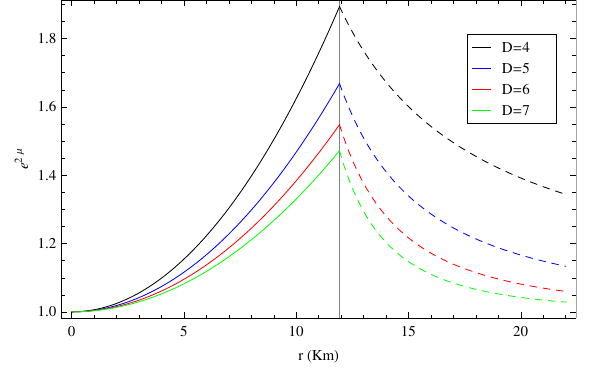}
		\caption{(a)}
	\end{subfigure}%
	\begin{subfigure}[t]{0.5\textwidth}
		\centering
		\includegraphics[width=1\textwidth]{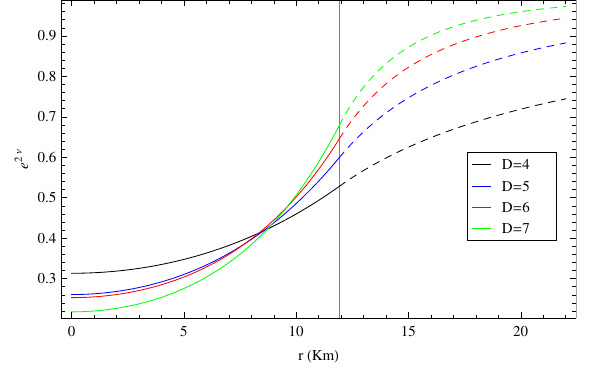}
		\caption{(b)}
	\end{subfigure}
	\caption{Radial variation of $e^{2\mu}$ (panel a) and $e^{2\nu}$ (panel b) inside PSR J1614-2230 in different space-time dimensions ($D$). The dashed lines correspond to exterior Schwarzschild solution. The vertical line represents boundary of the star. Here, the pressure anisotropy parameter $\alpha=0.3$ for $D=4,5$ and $0.98$ for $D=6,7$ respectively. Value of the spheroidal parameter $\lambda$ is set to 40.}
	\label{Fig5}
\end{figure}

In this section, we test the physical viability of our model to check whether it is consistent with realistic stellar structure. In order to fit the obtained model with observational mass radius data of compact object, we consider the millisecond pulsar PSR J1614-2230 having observed mass $1.908^{+0.016}_{-0.016}$ $M_{\odot}$ and radius $11.93^{+0.50}_{-0.50}$ $km$ \cite{Li}. In Figs.~\ref{Fig1}-\ref{Fig4}, we have graphically demonstrated the nature of energy density ($\rho$) and different pressures such as radial ($p_{r}$), tangential ($p_{t}$) and also the difference ($\Delta$).
In Fig.~\ref{Fig5}, the metric functions $e^{2\mu}$ and $e^{2\nu}$ have been plotted as a function of radial distance $'r'$ inside the compact object PSR J1614-2230. The corresponding exterior Schwarzschild solution have also been indicated in the figure. From Fig.~\ref{Fig5} we can infer that the interior solution is perfectly matched with the Schwarzschild exterior metric. Also, both the metric functions are found to be regular and singularity free inside the compact object as evident from the figures.
\subsection{Energy conditions}
\begin{figure}[ht!]
	\centering
	\includegraphics[width=0.5\textwidth]{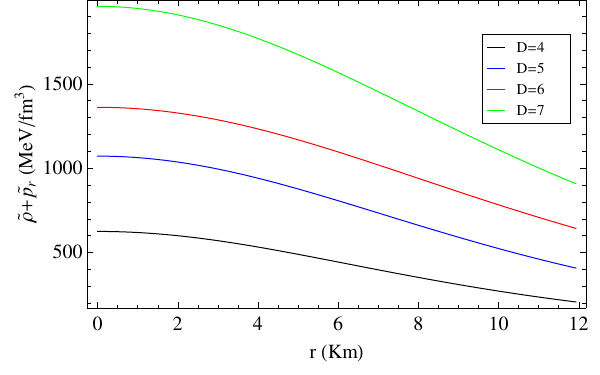}
	\caption{Radial variation of ($\tilde{\rho}+\tilde{p_{r}}$) inside PSR J1614-2230 in different space-time dimensions ($D$). Here, the pressure anisotropy parameter $\alpha=0.3$ for $D=4,5$ and $0.98$ for $D=6,7$ respectively. Value of the spheroidal parameter $\lambda$ is set to 40.}
	\label{Fig6}
\end{figure}

\begin{figure}[ht!]
	\centering
	\includegraphics[width=0.5\textwidth]{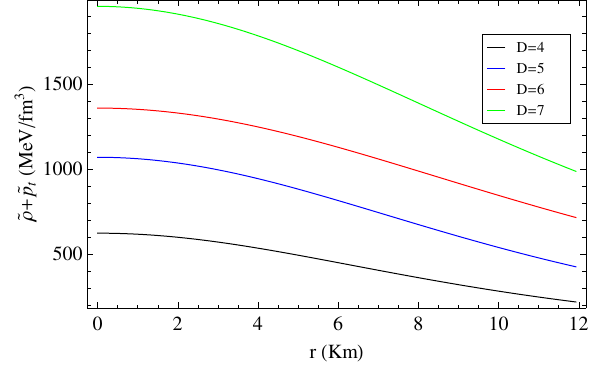}
	\caption{Radial variation of ($\tilde{\rho}+\tilde{p_{t}}$) inside PSR J1614-2230 in different space-time dimensions ($D$). Here, the pressure anisotropy parameter $\alpha=0.3$ for $D=4,5$ and $0.98$ for $D=6,7$ respectively. Value of the spheroidal parameter $\lambda$ is set to 40.}
	\label{Fig7}
\end{figure}

\begin{figure}[ht!]
	\centering
	\includegraphics[width=0.5\textwidth]{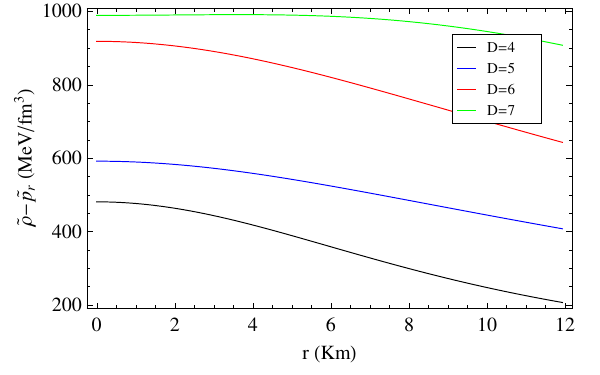}
	\caption{Radial variation of ($\tilde{\rho}-\tilde{p_{r}}$) inside PSR J1614-2230 in different space-time dimensions ($D$). Here, the pressure anisotropy parameter $\alpha=0.3$ for $D=4,5$ and $0.98$ for $D=6,7$ respectively. Value of the spheroidal parameter $\lambda$ is set to 40.}
	\label{Fig8}
\end{figure}

\begin{figure}[ht!]
	\centering
	\includegraphics[width=0.5\textwidth]{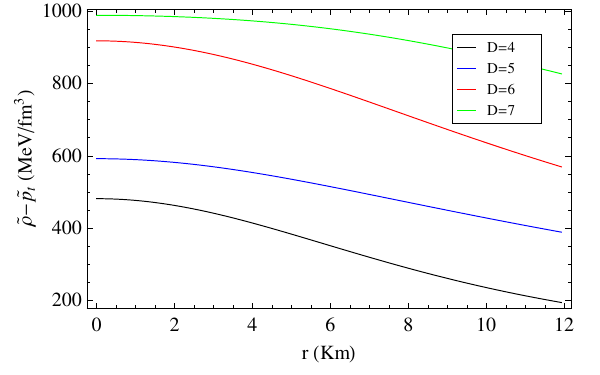}
	\caption{Radial variation of ($\tilde{\rho}-\tilde{p_{t}}$) inside PSR J1614-2230 in different space-time dimensions ($D$). Here, the pressure anisotropy parameter $\alpha=0.3$ for $D=4,5$ and $0.98$ for $D=6,7$ respectively. Value of the spheroidal parameter $\lambda$ is set to 40.}
	\label{Fig9}
\end{figure}

\begin{figure}[ht!]
	\centering
	\includegraphics[width=0.5\textwidth]{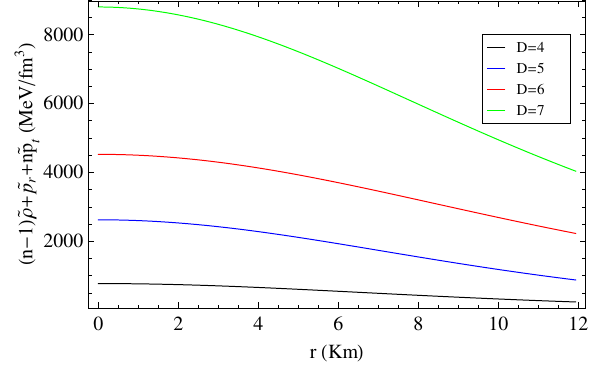}
	\caption{Radial variation of $(n-1)\tilde{\rho}+\tilde{p_{r}}+n\tilde{p_{t}}$ inside PSR J1614-2230 in different space-time dimensions ($D$). Here, the pressure anisotropy parameter $\alpha=0.3$ for $D=4,5$ and $0.98$ for $D=6,7$ respectively. Value of the spheroidal parameter $\lambda$ is set to 40.}
	\label{Fig10}
\end{figure}

In general relativity, curvature of space-time fabric influences matter to move and vice versa. This mutual interaction is encoded as a mathematical form in Einstein field equation. If one considers Einstein field equation without placing any constraints on the matter, any Lorentzian metric field on any manifold can be viewed as a valid solution. This can result in unexpected phenomena, including wormholes, faster-than-light travel, closed timelike curves, and other violations of causality. Since these phenomena have never been observed, an explanation is necessary. The most common explanation is that such spacetimes generally require matter fields with exotic properties, like negative energy densities. Energy conditions play a significant role in the context of these exotic properties. Energy conditions are pointwise restrictions on the energy-momentum tensor. In a $D$ dimensional space, energy conditions are essentially an eigen value problem and correspond to the root of a $D$-degree polynomial. Consequently, for a physically admissible solution of stellar structure, several energy conditions \cite{Brassel,Maeda} namely, (i) Null Energy Condition (NEC), (ii) Weak Energy Condition (WEC), (iii) Dominant Energy Condition (DEC), and (iv) Strong Energy Condition (SEC) should be fulfilled at all internal points. Mathematically, the energy conditions may be framed as follows:
\begin{itemize}
	\item {\bf NEC}: $\tilde{\rho}+\tilde{p_{r}}\geq 0$, $\tilde{\rho}+\tilde{p_{t}}\geq 0$.
	\item {\bf WEC}: $\tilde{\rho}\geq 0$, $\tilde{\rho}+\tilde{p_{r}}\geq 0$, $\tilde{\rho}+\tilde{p_{t}}\geq 0$.
	\item {\bf DEC}: $\tilde{\rho}\geq 0$, $\tilde{\rho}-\tilde{p_{r}}\geq 0$, $\tilde{\rho}-\tilde{p_{t}}\geq 0$.
	\item {\bf SEC}: $\tilde{\rho}+\tilde{p_{r}}\geq 0$, $\tilde{\rho}+\tilde{p_{t}}\geq 0$, $(n-1)\tilde{\rho}+\tilde{p_{r}}+n\tilde{p_{t}}\geq 0$. 
\end{itemize} 
In this work, we have studied all the energy conditions which are graphically visualised in Fig.~\ref{Fig1} and Figs.~\ref{Fig6}-\ref{Fig10}.

\subsection{Equation of state}
\begin{figure}[ht!]
	\centering
	\includegraphics[width=0.85\textwidth]{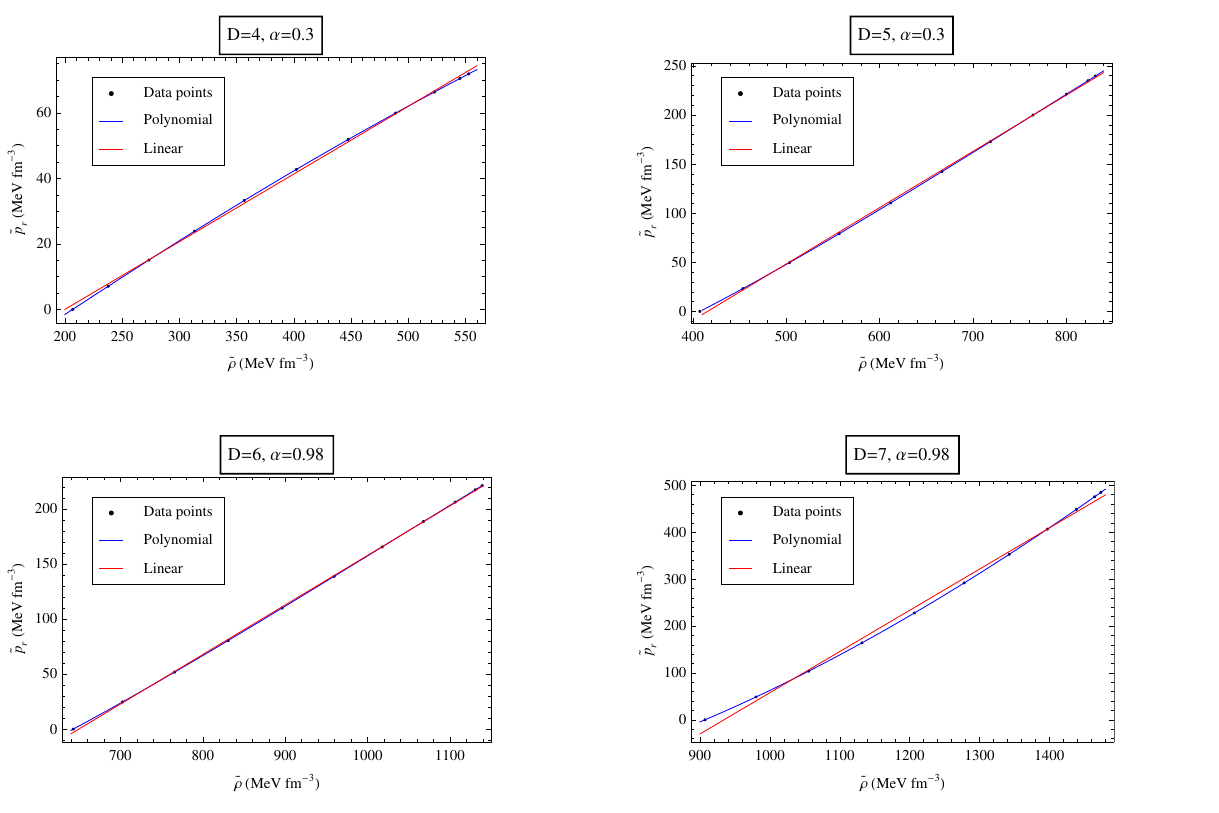}
	\caption{Equation of state for PSR J1614-2230 in different space-time dimensions ($D$).}
	\label{Fig11}
\end{figure}

The equation of state (EoS) defines the functional relation between pressure ($p_{r}$) and energy density ($\rho$) of a compact object. EoS plays a prominent role in defining characteristics of interior matter as well as mass radius relation of a compact object. In the present work, energy density ($\rho$) and pressure ($p_{r}$) are already obtained from the solution of the field equation~(\ref{eq3}) and are expressed in Eqs.~(\ref{eq12}) and (\ref{eq13}) respectively. However, decoding a functional relation between energy density ($\rho$) and pressure ($p_{r}$) is cumbersome owing to the complexity of the expressions of $\rho$ and $p_{r}$. Therefore, in our model, we adopt an alternative approach to find the relevant EoS of a compact stellar object. We at first assume a polynomial relation between pressure ($p_{r}$) and energy density ($\rho$) given by \cite{Chattopadhyay}
\begin{equation}
	p_{r}=\sum_{i=0}^{s}a_{i}\rho^{i}. \label{eq30}
\end{equation}  
Now, we use the observed mass, radius data of PSR J1614-2230 to find the values of $\rho$ and $p_{r}$ at various interior points of the compact object using Eqs.~(\ref{eq12}) and (\ref{eq13}). Then, we fit the obtained data with the predicted EoS given by Eq.~(\ref{eq30}). In Table~\ref{Tab2}, we have shown the coefficients $a_{i}$ estimated for the compact object PSR J1614-2230.
\begin{table}[ht!]
	\centering
	\caption{Values of the coefficients $a_{i}$ obtained from Eq.~(\ref{eq29}) for PSR J1614-2230.}
	\label{Tab2} 
	\begin{tabular}{ccccccc}
		\hline
		Dimension & \multicolumn{6}{c}{coefficients $a_{i}$}  \\
		($D$)  & $a_{0}$ & $a_{1}$ & $a_{2}$ & $a_{3}$ & $a_{4}$ & $a_{5}$ \\ \hline 
		4      & $-41.733$ & $0.141$   &$5.21\times10^{-4}$&$-1.4\times10^{-6}$&$1.6\times10^{-9}$&$-7.18\times10^{-13}$ \\
		5      & $-75.684$ & $-0.336$ &$19.93\times10^{-4}$&$-2.23\times10^{-6}$&$1.32\times10^{-9}$&$-3.28\times10^{-13}$ \\
		6      &$-67.966$ & $-0.451$ &$14.2\times10^{-4}$&$-1.12\times10^{-6}$&$0.45\times10^{-9}$&$-0.78\times10^{-13}$ \\
		7      & $-42.572$ & $-0.774$ &$12.41\times10^{-4}$&$-0.47\times10^{-6}$&$0.12\times10^{-9}$&$-0.15\times10^{-13}$ \\
		\hline
	\end{tabular}
\end{table}
Table~\ref{Tab2} shows that effect of the higher order coefficients are negligible. Accordingly, in our model we restrict our analysis up to $s=5$ only. In Fig.~\ref{Fig11}, we have shown the EoS of the interior matter of PSR J1614-2230 for linear and polynomial fitting of Eq.~(\ref{eq30}).
From Fig.~\ref{Fig11}, we observe that in each dimension, the polynomial fit is more accurate than the linear fit in our model. Therefore, any linear EoS such as MIT bag model EoS especially with a constant value of bag parameter cannot be used to analyse the possible accumulation of strange matter in a star using our model. However, as an alternative approach, we can treat the bag parameter ($B$) as a density dependent quantity ($B(\rho)$). In such a formalism, the pressure expression from the MIT bag EoS namely, $p_{r}=\frac{1}{3}(\rho-4B)$ can be eliminated using Eq.~(\ref{eq30}) so that we get a functional dependence of the bag parameter ($B$) on the density ($\rho$) as follows:
\begin{equation}
	B(\rho) = \sum_{i=0}^{l} k_i \rho^i, \label{eq31}
\end{equation}  
where the coefficients $k_i$ s are related to $a_i$ s through the relations $k_0=-\frac{3}{4}a_0$, $k_1=\frac{(1-3a_1)}{4}$, $k_j=-\frac{3}{4}a_j$, where $j$ runs from 2 to $s$.\par
In the MIT bag model, nucleon are considered as a 'bag' of perturbative vacuum and embedded in a non-perturbative or true vacuum. The quarks are assumed to be confined within the 'bag' under the action of a net inward force $B$ also known as bag parameter. With the rise of temperature, above the critical limit corresponds to the de-confinement quark phase. The difference in energy between the perturbative and non-perturbative vacua diminishes and consequently the net inward pressure ($B$) must vanish. This notion suggests that $B$ must be a temperature dependent quantity \cite{Muller}. In a similar manner, the de-confinement of quarks may also be realized by increasing the baryon number density above a threshold, which allows one to consider $B$ as a density dependent quantity \cite{Reinhardt}.\par
In the present work, it has been assumed that the temperature of a star is well below the typical interaction energy between the quark flavours. Therefore, the effect of temperature on $B$ has not been considered in the analysis. However, since the density has profound effects on gross properties of compact objects. Effect of density on $B$ cannot be ignored. In subsequent part of the article we shall be using Eq.~(\ref{eq31}) to study the possible stability of strange quark matter inside a compact object. The energy per baryon of strange matter reads as \cite{Weber}
\begin{equation}
	E_{B}=2\sqrt{3}\left(\frac{3\pi^{2}B(\rho)}{4}\right)^{\frac{1}{4}}. \label{eq32}
\end{equation}
\begin{figure}[ht!]
	\centering
	\includegraphics[width=0.85\textwidth]{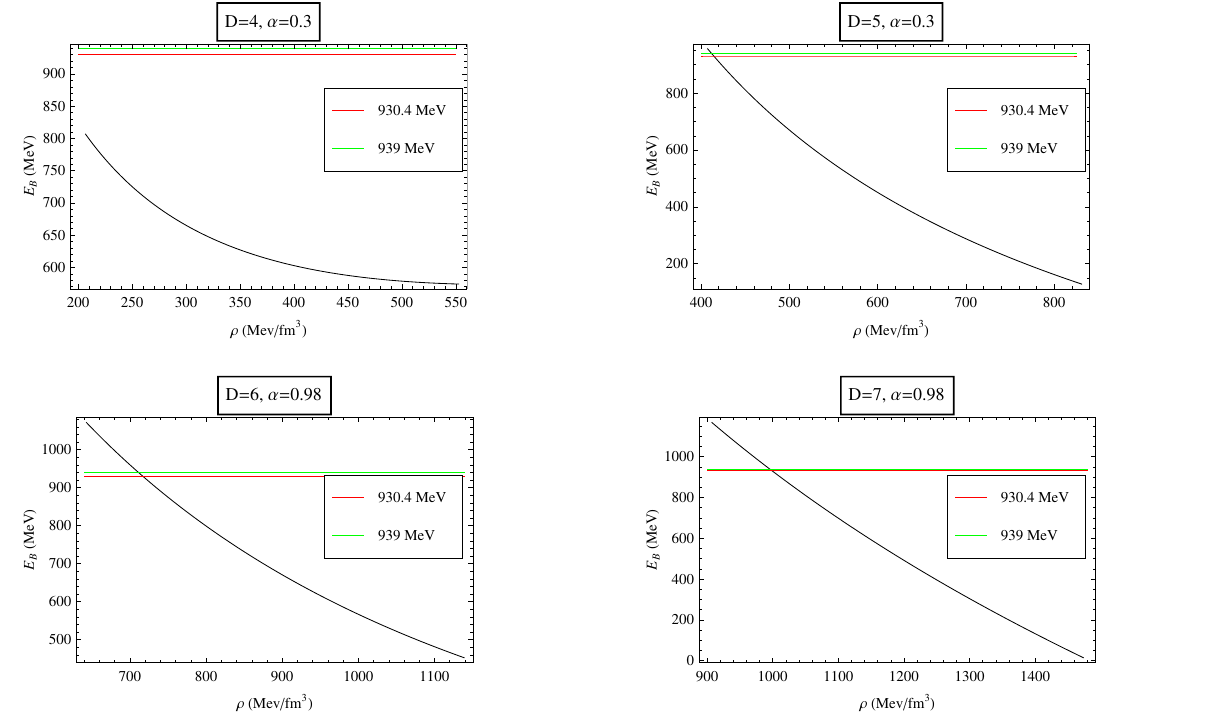}
	\caption{Variation of energy per baryon ($E_{B}$) of 3-flavour quark matter with density ($\rho$) of star for PSR J1614-2230 in different space-time dimensions ($D$). The red and green lines represent binding energy and mass correspond to $^{56}Fe$ ($930.4$ $MeV$) and nucleon ($939$ $MeV$) respectively.}
	\label{Fig12}
\end{figure}
Fig.~\ref{Fig12} shows the variation of energy per baryon ($E_{B}$) with density of PSR J1614-2230 in different space-time dimensions ($D$). It is observed that in four dimensions, strange matter is absolutely stable relative to $^{56}Fe$ ($E_{B}<930.4$ $MeV$). However, with increase in the dimension of space-time, the stability window shifts towards metastability ($939 MeV$ $<E_{B}<$ $930.4 MeV$) and instability ($E_{B}>939$ $MeV$).
\begin{figure}[ht!]
	\centering
	\includegraphics[width=0.85\textwidth]{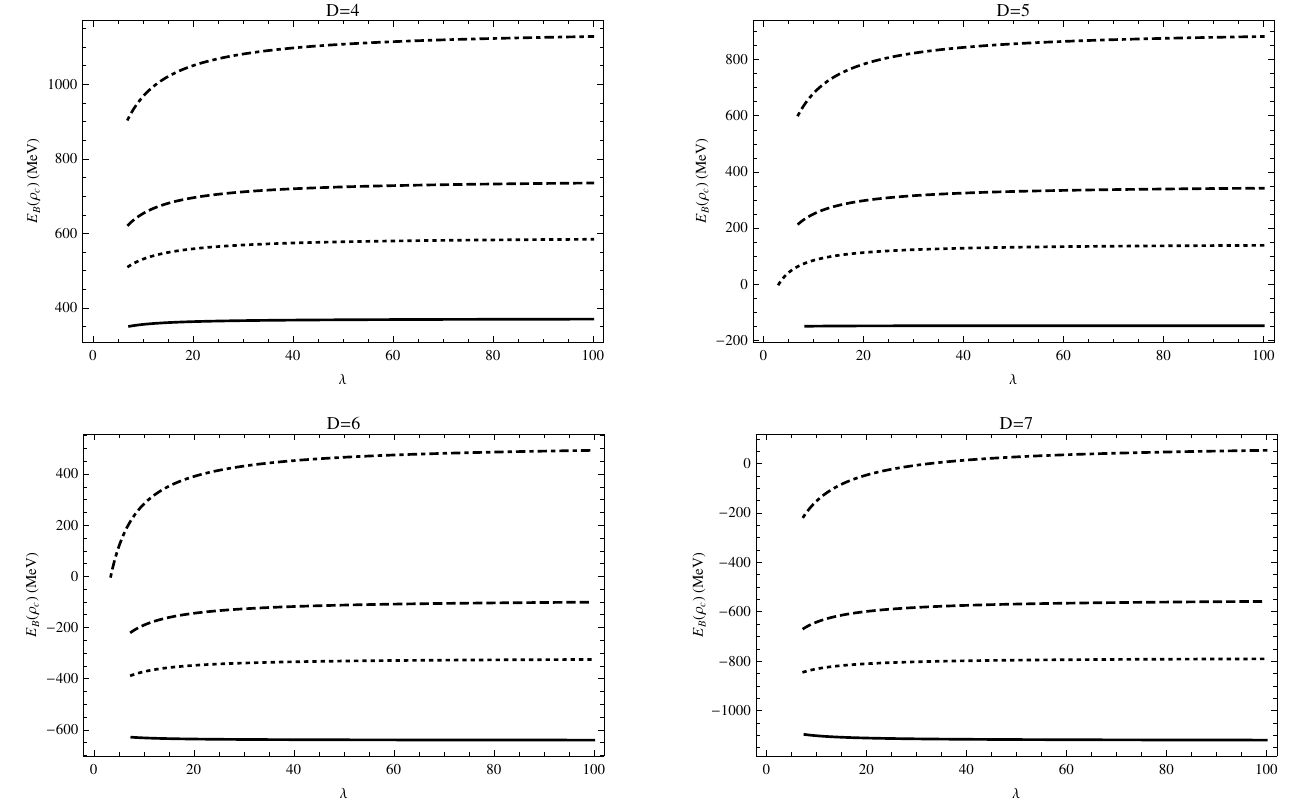}
	\caption{Variation of central energy per baryon ($E_{B}(\rho_{c})$) of 3-flavour quark matter with spheroidal parameter ($\lambda$) for PSR J1614-2230 in different space-time dimensions ($D$). In each dimension plots from bottom to top correspond to $\alpha=$ 0, 0.3, 0.5 and 0.98 respectively.}
	\label{Fig13}
\end{figure}
In Fig.~\ref{Fig13}, we have shown the dependence of central value of energy per baryon ($E_{B}(\rho_{c})$) on the spheroidal parameter ($\lambda$) of the star PSR J1614-2230. It is apparent from the Fig.~\ref{Fig12}, that both dimension ($D$) and pressure anisotropy ($\alpha$) have significant role on the overall stability of SQM. When $D>4$, higher amount of anisotropy is required otherwise $E_{B}$ becomes negative which is physically absurd. Therefore, it may be concluded that space-time dimension and anisotropy are inter related and at higher dimensions, a star might switch from isotropic to anisotropic as far as stability of its quark matter content is concerned. For better visualization of the problem, we have also plotted $E_{B}(\rho_{c}$) against pressure anisotropy parameter ($\alpha$) in Fig.~\ref{Fig14}.
\begin{figure}[ht!]
	\centering
	\includegraphics[width=0.7\textwidth]{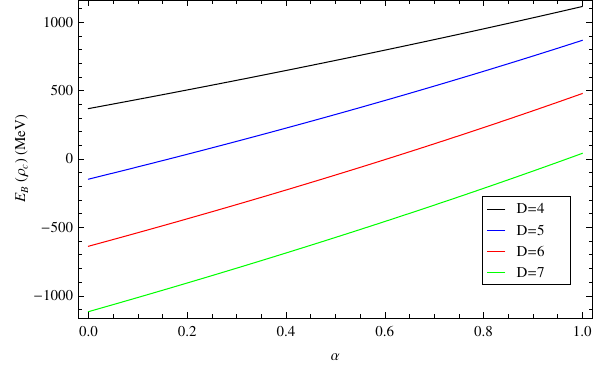}
	\caption{Variation of central energy per baryon ($E_{B}(\rho_{c})$) of 3-flavour quark matter with pressure anisotropy parameter ($\alpha$) for PSR J1614-2230 in different space-time dimensions ($D$). To obtain the plots $\lambda$ has been set to 40.}
	\label{Fig14}
\end{figure}  
Fig.~\ref{Fig14} shows that for $D>4$, there exists a critical value of $\alpha$ ($\alpha_{crit}$) below which $E_{B}(\rho_{c})<0$. From the figure we note that $\alpha_{crit}$ $\approx 0.16$ when $D=5$, $\approx 0.61$ when $D=6$, and $\approx 0.97$ when $D=7$ respectively.
\section{Mass-radius relation}\label{mass}
\begin{figure}[ht!]
	\centering
	\includegraphics[width=0.85\textwidth]{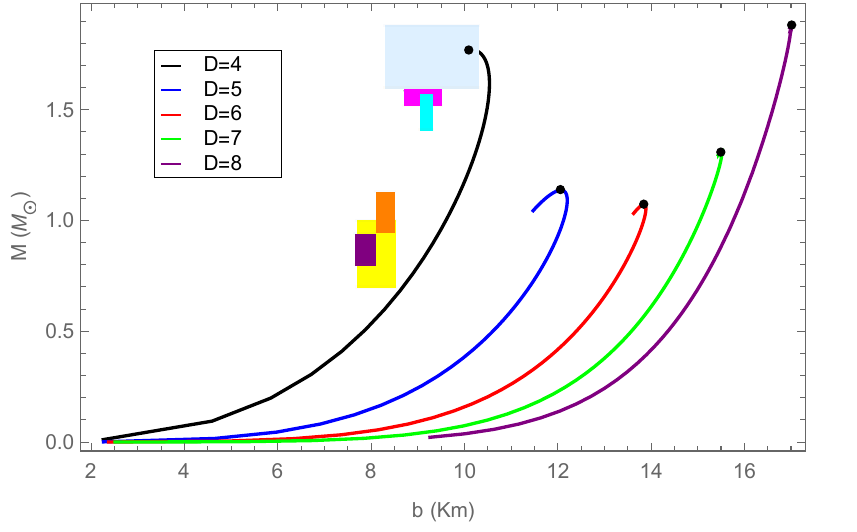}
	\caption{Variation of mass ($M$) of compact objects with radius ($b$) for different values of $D$. Here the shaded regions correspond to the observed masses and radii of some compact objects such as, Yellow region: Her X-1, Magenta region: 4U 1820-30, Cyan region: Cen X-3, Orange region: LMC X-4, Purple region: 4U 1538-52, Light-blue region: 4U 1608-52.}
	\label{Fig15}
\end{figure}

\begin{figure}[ht!]
	\centering
	\includegraphics[width=0.85\textwidth]{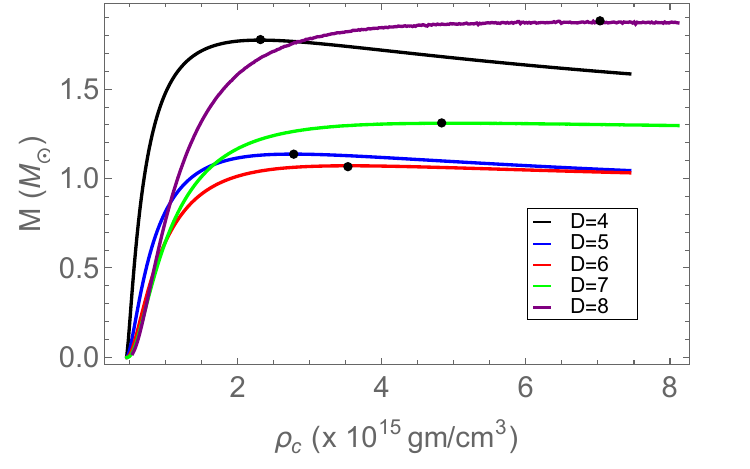}
	\caption{Variation of mass ($M$) of compact objects with central density ($\rho_{c}$) for different values of $D$.}
	\label{Fig16}
\end{figure}

\begin{figure}[ht!]
	\centering
	\includegraphics[width=0.8\textwidth]{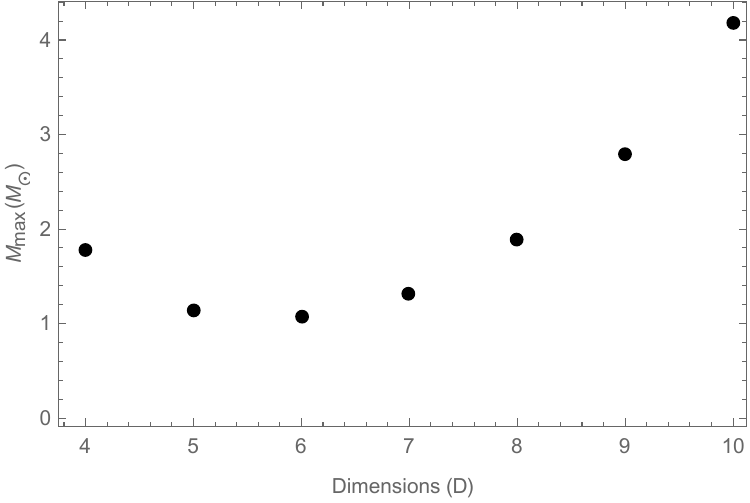}
	\caption{Variation of maximum mass ($M_{max}$) of compact objects with space-time dimensions ($D$).}
	\label{Fig17}
\end{figure}

The relation between mass and radius of compact objects is a very important feature which allows one to distinguish among various types of compact objects. In the present work, we have obtained the mass-radius relation by plotting the total mass ($M$) of compact objects within its radius $b$ by solving the TOV equations in $D$-dimensional space \cite{Bakhti}. To solve the TOV equation, we have made use of the EoS given in Eq.~(\ref{eq30}). The result we have obtained is shown in Fig.~\ref{Fig15}.
In Fig.~\ref{Fig16}, variation of mass ($M$) of has been studied as a function of central density ($\rho_{c}$) of compact objects. It has been found that below the maximum mass points, mass increases with central density thus indicating that $(\frac{\partial M}{\partial \rho_c})>0$. Therefore, our model satisfies the Harrison-Zeldovich-Novikov's static stability criteria \cite{Zeldovich,Harrison}. 
The maximum mass, radius and compactness thus obtained are listed in Table~\ref{Tab3}\footnote{The total mass ($M$) of a star obtained from TOV equation has the dimension of $mass\times length^{D-4}$. However, the physical mass of a star has the dimension of mass and is equal to $\frac{M}{b^{D-4}}$ which is finally converted into solar mass unit for plotting.}. The Buchdahl limit in $D$-dimensions is also indicated in the table.
\begin{table}[ht!]
	\centering
	\caption{Maximum value of stellar parameters in different dimensions ($D$).}
	\label{Tab3} 
	\begin{tabular}{cccccc}
		\hline
		Dimension & Max. mass & Max. radius & Max. central density & Max. compactness & Buchdahl bound \\
		$D$ & ($M_{max}(M_{\odot})$)& $b_{max}(Km)$& $\rho_{c_{max}}\times10^{15}(gm/cm^3)$& $\frac{G}{c^2} \frac{M_{max}}{b_{max}^{D-3}}$ & $2\frac{D-2}{(D-1)^2}$\cite{Ponce} \\\hline 
		4   & 1.77 & 10.09 & 2.30 & $2.60\times10^{-1}$ & 0.444\\
		5   & 1.14 & 12.07 & 2.77 & $1.15\times10^{-2}$ & 0.375\\
		6   & 1.07 & 13.84 & 3.53 & $0.59\times10^{-3}$ & 0.320\\
		7   & 1.31 & 15.50 & 4.83 & $0.33\times10^{-4}$ & 0.277\\
		8   & 1.88 & 16.99 & 7.02 & $1.96\times10^{-6}$ & 0.245\\
		9   & 2.80 & 18.18 & 7.01 & $1.14\times10^{-7}$ & 0.218\\
		10  & 4.19 & 19.05 & 7.02 & $6.78\times10^{-9}$ & 0.197\\
		11  & 6.30 & 19.82 & 7.02 & $3.90\times10^{-10}$ & 0.180\\
		\hline
	\end{tabular}
\end{table}
From Table~\ref{Tab3}, it is clearly noticeable that the Buchdahl bound is sustained up to $D=11$ in our model, and from this obtained trend, we may assert that Buchdahl limit will be satisfied for $D>11$ also. The present framework with the specific EoS in Eq.~(\ref{eq30}) or equivalently, the density dependent bag parameter $B(\rho)$ in Eq.~(\ref{eq31}) provides the flexibility to allow for such higher dimensions $(D>11)$. To visualize the dependence of maximum mass on dimension $D$, we have plotted the maximum mass against space-time dimension $D$ in Fig.~\ref{Fig17} which shows that the maximum mass at first decreases with $D$, takes a minimum value when $D=6$ then increases rapidly. It appears that the influence of mass on compactness is greater than the radius until $D=6$ then the effect reverses i.e. the effect of radius dominates over mass for $D>6$.
\section{Stability Analysis}\label{stab}
We have checked the stability of our model from the following points of view:\\
(i) Generalised TOV equation,\\
(ii) Herrera cracking condition, and\\
(iii) Variation of adiabatic index.
\subsection{Generalised TOV equation}
\begin{figure}[ht!]
	\centering
	\includegraphics[width=0.6\textwidth]{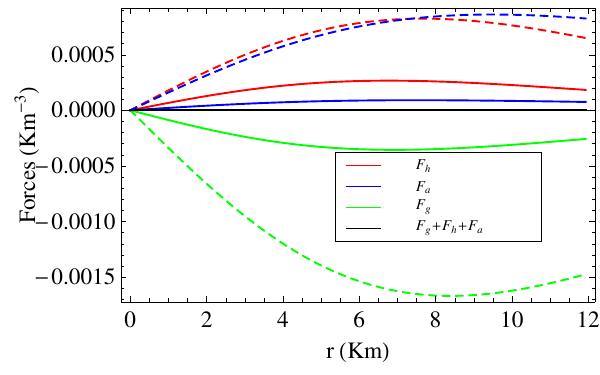}
	\caption{Variation of different forces inside PSR J1614-2230 for $D=4$ (solid curves) and $D=6$ (dashed curves) respectively.}
	\label{Fig18}
\end{figure}

In GR, the generalised TOV equation describes the hydrostatic equilibrium condition of a stellar object \cite{Tolman,Oppenheimer}. Mathematically, the condition is represented as 
\begin{equation}
	-\frac{M_{G}(\rho+p_r)}{r^2}e^{\lambda-\nu}-\frac{dp_r}{dr}+\frac{n}{r}(p_t-p_r)=0, \label{eq33}
\end{equation}
where the active gravitational mass $M_{G}$ within a radius $r$ can be obtained from Tolman-Whittaker formula \cite{Gron} as:
\begin{equation}
	M_G=r^2 e^{(\nu-\lambda)}\nu^\prime. \label{eq34}
\end{equation}
Substituting Eq.~(\ref{eq34}) into Eq.~(\ref{eq33}), we get
\begin{equation}
	-\nu^\prime(\rho+p_r)-\frac{dp_r}{dr}+\frac{n}{r}\Delta=0, \label{eq35}
\end{equation}
where $\Delta=p_t-p_r$. In Eq.~(\ref{eq35}), the first term represents gravitational force ($F_g$), second term hydrostatic force ($F_h$), and the last term is known as anisotropic force ($F_a$). In terms of these three forces Eq.~(\ref{eq35}) can be written as
\begin{equation}
	F_{g}+F_{h}+F_{a}=0. \label{eq36}
\end{equation} 
The above mentioned forces are shown graphically in Fig.~\ref{Fig18} for the compact object PSR J1614-2230, which shows that the condition of hydrostatic equilibrium is maintained throughout the interior of the compact object as the sum of all the forces vanishes.
\subsection{Herrera cracking condition}
In relativistic astrophysics, it is crucial that anisotropic stellar models remain stable against fluctuations in their physical parameters. In this regard, Herrera \cite{Herrera2} introduced the concept of 'cracking', which assesses the stability of an anisotropic matter distribution. Building on Herrera's idea, Abreu \cite{Abreu} developed a criterion stating that a stellar model is stable if the radial ($v_{r}$) and tangential ($v_{t}$) sound velocities satisfy the specified condition given below:
\begin{equation}
	0\leq |v_{t}^2-v_{r}^2|\leq 1. \label{eq37}
\end{equation}
\begin{figure}[ht!]
	\centering
	\includegraphics[width=0.6\textwidth]{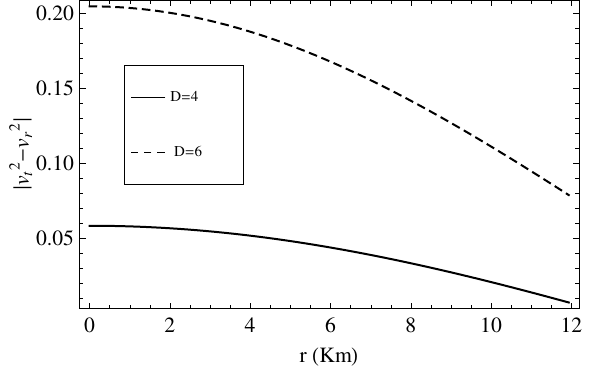}
	\caption{Radial variation of $|v_{t}^2-v_{r}^2|$ inside PSR J1614-2230 for $D=4$ and 6 respectively.}
	\label{Fig19}
\end{figure}
In Fig.~\ref{Fig19}, we have shown the variation of $|v_{t}^2-v_{r}^2|$ inside the compact object PSR J1614-2230 and it is evident that Abreu's inequality given by Eq.~(\ref{eq37}) is maintained in our model.
\subsection{Adiabatic index}
\begin{figure}[ht!]
	\centering
	\includegraphics[width=0.6\textwidth]{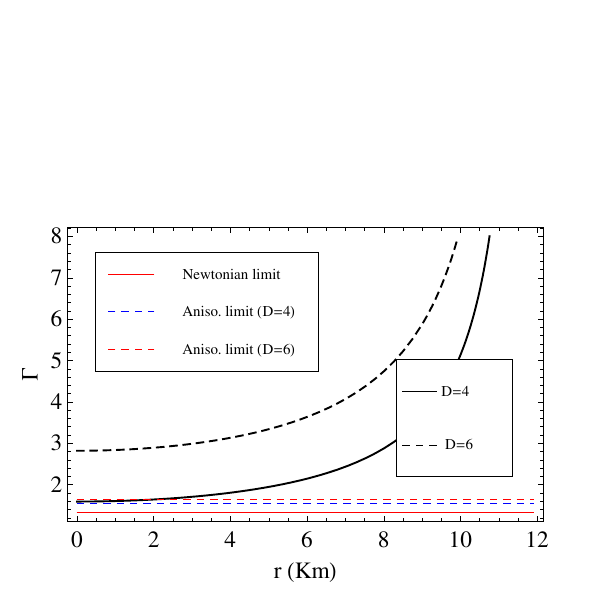}
	\caption{Radial variation of $|v_{t}^2-v_{r}^2|$ inside PSR J1614-2230 for $D=4$ and 6 respectively.}
	\label{Fig20}
\end{figure}

The adiabatic index ($\Gamma$) acts as a basic ingredient of the condition of dynamical stability postulated by  Chandrasekhar \cite{Chandrasekhar}. The stability of a compact objects manifests as an interconnection between the EoS describing the interior fluid distribution and the strength of the relativistic field. Consequently, adiabatic index forms a bridge between these two factors. Mathematically, the adiabatic index for isotropic star is defined as
\begin{equation}
	\Gamma=\frac{\rho+p_{r}}{p_{r}}\left(\frac{dp_{r}}{d\rho}\right). \label{eq38}
\end{equation} 
According to Heintzmann and Hillebrandt \cite{Heintzmann}, a Newtonian isotropic fluid sphere will be in stable equilibrium if $\Gamma>4/3$ and in neutral equilibrium if $\Gamma=4/3$. However, for an anisotropic star, Chan et al. \cite{Chan} modified the condition of stability, and accordingly we have:
\begin{equation}
	\Gamma>\gamma, \label{eq39}
\end{equation}
where, $\gamma=\frac{4}{3}-\left[\frac{4(p_r-p_t)}{3|p_r^{\prime}|r}\right]_{max}$. The variation of $\Gamma$ inside PSR J1614-2230 is depicted in Fig.~\ref{Fig20}. It may be noticed from the figure that $\Gamma$ is always greater than the Newtonian limit $4/3$ as well as the anisotropic limit ($\gamma$) given by Chan et al. indicating that the model is stable.
\section{Conclusions}\label{conc}
In this article, we have presented a method of generating exact solution to the Einstein field equation for a spherically symmetric perfect fluid distribution taking local anisotropy in pressure into account in the context of general relativity. To simplify the inherent non-linearity of the field equation, a specific form of the $g_{rr}$ metric potential has been assumed in accordance with the proposed concept of Vaidya and Tikekar \cite{Tikekar}. The solution thus obtained for the Vaidya-Tikekar metric ansatz and a specific form of anisotropic pressure is found to be regular and non-singular and capable of describing relativistic compact objects. Vaidya-Tikekar ansatz is specifically suitable for superdense stellar model. In $D$ dimensions, this particular ansatz represents the geometry of a $(D-1)$-spheroid when immersed in a $D$ dimensional Euclidean space. The metric ansatz is specified by a spheroidal parameter($\lambda$) and a geometry dependent radial parameter $R$. The obtained solution is basically a higher dimensional anisotropic extension of the solution obtained earlier by Mukherjee et al. \cite{Mukherjee}. The causality bound of the radial sound velocity ($v_{r}$) puts a restriction on the lower limit of the spheroidal parameter ($\lambda$) and depends both on the anisotropy parameter ($\alpha$) and space-time dimensions ($D$). It has been found that in isotropic case ($\alpha=0$) if we consider only four dimensions ($D=4$), we recover the bound $\lambda>\frac{3}{17}$ previously obtained by Mukherjee et al. \cite{Mukherjee}. It is interesting to note that the effect of $\alpha$ on the lower limit of $\lambda$ becomes significant when the number of space-time dimensions increases. To show physical analysis of our solution, we have considered the compact star PSR J1614-2230. For the observed value of mass and radius of the compact object mentioned above, we have drawn all the plots such as energy density ($\tilde{\rho}$), radial ($\tilde{p_{r}}$) and tangential ($\tilde{p_{t}}$) pressures and pressure anisotropy ($\tilde{\Delta}$) and are shown in Figs.~\ref{Fig1}-\ref{Fig4} respectively. The behaviours of the metric potentials ($e^{2\mu}$ and $e^{2\nu}$) are graphically illustrated in Fig.~\ref{Fig5}. From Figs.~\ref{Fig1}-\ref{Fig5}, it is clear that all the physical parameters derived from the solution of EFE are regular and non-singular and therefore relevant for the correct description of ultra-dense compact objects. The energy conditions are shown in graphical form in Fig.~\ref{Fig1} and Figs.~\ref{Fig6}-\ref{Fig10} and indicate that all the energy conditions are obeyed in the present model. To study the behaviour of matter inside a compact object, it is necessary to obtain a functional relation between pressure ($p_{r}$) and energy density ($\rho$) which is generally termed as the equation of state of the matter composed of the compact object. In the present analysis due to complexity of the mathematical expressions of $p_{r}$ and $\rho$, it is very difficult to arrive at any functional relationship between the two. Therefore, in this article we have adopted a different technique. We have at first, assumed a polynomial relation between $p_{r}$ and $\rho$ and from the observed mass and radius of PSR J1614-2230, evaluated a set of data of $p_{r}$ and $\rho$ at various interior points. Next, we have fitted the obtained data set with the proposed EoS and the related coefficients are tabulated in Table~\ref{Tab2}. The EoS for both linear and quadratic fits have shown in Fig.~\ref{Fig11} which signifies that in each dimension, the polynomial fitted EoS matches well with the observed data set. Therefore, it may be concluded that the present solution of relativistic compact objects with Vaidya-Tikekar ansatz does not represent matter distribution with linear form of EoS. This in turn indicates that MIT bag model type EoS with constant $B$ can not be associated with our model. However, if we consider bag parameter ($B$) as a density dependent quantity (as in Eq.~(\ref{eq31})) rather than a constant, then we get a modified form of the MIT bag EoS which resembles as a polynomial EoS. Therefore, the present analysis is still capable of admitting MIT bag EoS if $B$ is taken as a function of density ($\rho$). Considering density dependent bag model, energy per baryon ($E_{B}$) of SQM has been evaluated and the variation has been studied as a function of energy density ($\rho$) as displayed in Fig.~\ref{Fig12} and it can be noticed that with increased value of space-time dimensions ($D$), the stability window shifts towards instability through metastability. Since EFE represents a connection between matter and geometry, stability of quarks inside a compact star might be closely related to the associated space-time geometry. Therefore, it is customary to discuss the dependence of $E_{B}$ on some geometrical parameters such as $\lambda$. For this purpose, we have visualised the variation of $E_{B}$ with spheroidal parameter $\lambda$ in Fig.~\ref{Fig13} for different value of anisotropy parameter ($\alpha$). Fig.~\ref{Fig13} shows that value of $E_{B}$ rises with the increase of anisotropy inside a star. Interestingly, for $D>4$, we observe that a critical value of anisotropy ($\alpha_{crit}$) exists below which $E_{B}$ becomes negative which is physically absurd. For better insight of the relation between $E_{B}$ and $\alpha$, we have studied the variation of central value of $E_{B}$ ($E_{B}(\rho_{c})$) with $\alpha$ graphically in Fig.~\ref{Fig14} which indicates that the critical value of $\alpha$ increases with increase of dimensions ($D$), e.g. when $D=5$, $\alpha_{crit}\approx0.16$, for $D=6$, $\alpha_{crit}\approx0.61$ and for $D=7$, $\alpha_{crit}\approx0.97$ respectively. In Fig.~\ref{Fig15}, mass-radius relation of compact object has been studied by solving TOV equation for the prescribed EoS in Eq.~(\ref{eq30}). Fig.~\ref{Fig15} shows that the obtained mass-radius range covers a wider band of compact objects such as Her X-1, 4U 1820-30, Cen X-3, LMC X-4, 4U 1538-52, and 4U 1608-52. The maximum mass obtained from the plot have been registered in Table~\ref{Tab3} which shows that Buchdahl bound in $D$ dimensions \cite{Ponce} is obeyed up to $D=11$, and the results indicate that within this parameter space, the Buchdahl bound will be sustained, even for $D>11$ also. Therefore, as $M$-theory suggests maximum allowable space-time dimensions to be 11 \cite{Schwarz} and brane world cosmology allows any number of space-time dimensions, in our analysis it appears that relativistic stellar astrophysics is also at par with such permitted high values of space-time dimensions owing to the flexibility of the model. In Fig.~\ref{Fig16}, mass ($M$) obtained from the solution of TOV equation have been plotted against central density ($\rho_{c}$) of compact objects. It can be noted that up to the maximum mass points, $\left(\frac{\partial M}{\partial \rho_{c}}\right)>0$ and this condition signifies stable stellar configuration according to Harrison \cite{Harrison}, Zeldovich and Novikov \cite{Zeldovich}. To find dimensional dependency of maximum mass ($M_{max}$), a graphical representation has been made in Fig.~\ref{Fig17}. The graphical nature suggests that maximum mass decreases with dimensions, takes a minimum value at $D=6$, then rises with dimensions. Figs.~\ref{Fig18}-\ref{Fig20} depict the stability of stellar system in view of balancing of various forces, Cracking criterion given by Herrera \cite{Herrera2,Abreu}, and variation of adiabatic index. All of these figures reflect the stability of our model. 
\section{Acknowledgement}\label{ack}
KBG is grateful to IUCAA, Pune, India for extending their visiting facilities. DB is thankful to DST, Govt. of India for providing the fellowship vide no. DST/INSPIRE Fellowship/2021/IF210761. PKC acknowledges, with gratitude, the support rendered by IUCAA, Pune, India through the visiting associateship program, where the work has been completed. 
\begin{description}
	\item[Availability of data and material:] 
	This manuscript has no associated data or the data will not be deposited, we have used only observed mass and radius of some known compact objects to construct relativistic stellar models.
\end{description}
\bibliographystyle{elsarticle-num}
	
\end{document}